# Universal diamond edge Raman scale to 0.5 terapascal: The implication to metallization of hydrogen.

M. I. Eremets[1]*, V. S. Minkov[1], P. P. Kong[1], A. P. Drozdov[1], S. Chariton[2], V. B. Prakapenka[2]

[1]*Max Planck Institute for Chemistry; Hahn Meitner Weg 1, Mainz, 55128, Germany*
[2]*Center for Advanced Radiation Sources, University of Chicago; 5640 South Ellis Avenue, Chicago, Illinois, 60637, USA*

*Corresponding author. Email: m.eremets@mpic.de

**Abstract**

The recent progress in generating static pressures up to terapascal values opens opportunities for studying novel materials with unusual properties, such as metallization of hydrogen and high-temperature superconductivity. However, an evaluation of pressure above ~0.3 terapascal is a challenge. We report a universal high-pressure scale up to ~0.5 terapascal, based on the shift of the Raman edge of stressed diamond anvils correlated with the equation of state of Au and does not require an additional pressure sensor. According to the new scale, the pressure values are substantially lower by 20% at ~0.5 terapascal compared to the extrapolation of the existing scales. We compared the available data of $H_2$ at the highest static pressures. We showed that the onset of the proposed metallization of molecular hydrogen reported by different groups is consistent when corrected with the new scale and can be compared with various theoretical predictions.

The terapascal pressure range has recently been achieved in dynamic compression experiments at large facilities (*1*). Amazingly, similar pressures can be generated in static experiments using the incommensurably smaller and simpler device – a diamond anvil cell (DAC)(*2, 3*), appropriate for many more diversified studies. Under such extreme pressure conditions, the structure and properties can change drastically even at ambient temperature because the work associated with the *P-V* compression is comparable to the energy of atomic bonds. For example, the "simple metal" sodium transforms to an insulator (electride)(*4*); conversely, oxygen and xenon transform to metals(*5, 6*). Molecular nitrogen transforms into an atomic state with a diamond-like structure(*7*).

Motivation for generating multi-megabar static pressures is primarily driven by the pursuit of obtaining atomic metallic phase of hydrogen, which is predicted to be a room temperature superconductor at pressures of ~500 GPa and may combine superconductivity and superfluidity(*8*). There has been substantial progress in the study of hydrogen, both theoretically and experimentally. Electrical conductivity studies showed that molecular hydrogen metalizes at ~320 GPa through the closing of the indirect gap(*9-11*). At higher pressures, a direct bandgap likely closes according to the abrupt decrease in the infrared absorption at ~427 GPa (*12*) and disappearance of the Raman modes at ~450 GPa(*10*). The problem of metallic hydrogen, in turn, has stimulated the search for high-temperature superconductivity in hydrogen-rich compounds. Nearly room temperature superconductivity with a $T_c$ ~ of 203 K in $H_3S$(*13*) and $T_c$ ~ 250 K in $LaH_{10}$ (*14, 15*) was reached at pressures of ~150-170 GPa. The latest theoretical calculations predict $T_c$s exceeding room temperature: $T_c$ ~330 K in $CeH_{18}$ (*16*), and ~470 K for $Li_2MgH_{16}$ (*17*) at higher pressures of ~300-500 GPa.

Static pressures up to 1 TPa were gradually increased starting from the modest ~10 GPa by improving the diamond anvil geometry and DAC design. A decisive step was the development of beveled anvils(*18*), which instantly opened the way for pressures beyond 100 GPa. With more complicated profiling of anvils, such as toroidal grooves, which surpass the plain Bridgeman anvils(*19,*

20) pressures of ~400 GPa(*2, 12*) up to a maximum of ~600 GPa(*2*) were achieved. Encouragingly, ~1 TPa was reached between two tiny hemispherical nanocrystalline diamonds interposed between conventional diamond anvils(*3*).

However, determining the pressure inside a DAC is difficult. The absolute values of pressure can be derived from a combination of *V(P)* data measured by X-ray diffraction and derivative d*V*/d*P* data measured by ultrasonic or Brillouin scattering, but it is limited to 120 GPa(*21*). The pressure estimation in DACs has been based on temperature-corrected shockwave compression data(*22*). In contrast to adiabatic compression in shock waves, the recently developed shockless ramp technique provides nearly isentropic conditions and lower heating effects. According to Ref. (*23*) the pressure along the isentrope at 600 GPa differs by only ~6 GPa compared to the 298 K isotherm. The isotherms of copper, gold, platinum and other materials(*23*) can be used for pressure estimation in DACs; however, a synchrotron X-ray source is needed to probe the lattice volume of the standards.

Instead, more convenient secondary scales are widely used, such as the ruby luminescence scale calibrated to 156 GPa against the equation of state of metals(*24*). However, the applicability of the ruby scale is limited to ~200 GPa because of the drastic weakening of the ruby luminescence(*25*).

The only practical alternative to X-ray probes for pressure determination above ~200 GPa is the diamond edge Raman scale. It is based on the fact that a Raman spectrum measured from a sample in a DAC inevitably contains a signal from the stressed anvils – a strong, broad band with a well-defined cutoff(*25, 26*). This high-wavenumber edge correlates with the pressure in the sample. The band at lower wavenumbers stems from the Raman signal of the diamond at deeper regions, where stresses are lower. Hanfland and Syassen(*26*) analyzed the Raman spectra of stressed anvils and proposed a linear pressure dependence of the diamond Raman edge up to 30 GPa. A linear pressure dependence with different constants was also suggested up to ~200 GPa(*27*). However, there are uncertainties that these pressure dependencies can be used as a practical scale with sufficiently high precision. As noted in Ref.(*26*), the stress pattern within the strained diamond anvil differs from the considered uniaxial case, which is highly anisotropic and possibly depends on the geometry of the diamond anvils and gasket material, pressure medium, and/or sample stiffness.

The actual diamond pressure scale was established based on numerous experiments at pressures up to 200 GPa with a ruby chip and X-ray pressure sensors(*9, 25*). It was found that the nonlinear pressure-induced shift of the diamond Raman edge is surprisingly suitable for a reliable estimation of pressure in diamond anvils of different shapes, with various gaskets and samples(*25*). Akahama and Kawamura(*28, 29*) extended the diamond pressure scale to 410 GPa by correlating the diamond Raman edge with the X-ray diffraction data of a Pt sample. However, the pressure estimation above ~250 GPa has since remained uncertain, particularly because the extended scale(*29*) apparently deviates from the extrapolation of the well-established low-pressure data(*25, 28*) reaching ~10% at ~400 GPa. The authors(*29*) suggested that the reason for the puzzling deviation might be a certain change in the stress state on the culet face of diamond anvils or the accuracy of the equation of state of Pt used (*30*). However, the updated equation of state of Pt did not overcome this contradiction(*23*). The assumptions that the extended scale(*29*) overestimates pressure values were made based on abnormal deviation of the loading curve(*12*) and the pressure dependence of the frequency of hydrogen vibron above ~300 GPa(*12, 31*). The estimation of pressure above 400 GPa is even more uncertain because it is based only on extrapolations of different scales, which yield unacceptably large deviations up to ~100 GPa at 500 GPa(*9, 25, 28, 29*).

The present study establishes a reliable diamond-edge Raman scale at pressures up to ~500 GPa. The accurate determination of pressure values in the 250-500 GPa pressure range is crucial because of the recent significant progress in the generation of extremely high static pressures, the study of metallic hydrogen and high-temperature superconductivity, phase transitions, chemistry, and the necessity of comparison between experiments of different groups and theoretical predictions, which are becoming more and more precise.

We prepared eight DACs with a gold sample as the X-ray pressure standard. The single-crystal diamond anvils had a culet plane normal to the [001] crystallographic direction. We created profiles of anvil tips of different shapes to achieve the highest pressure (Fig. S1). We established the correlation between the lattice parameters of a gold sample probed with X-ray diffraction and the high-wavenumber Raman edge of stressed diamond anvils from the same point of the sample (Fig. 1). The equation of state of gold was taken from the isotherm derived from the ramp data(*23*) (see Supplementary Materials for a detailed description of the experiment). We plotted this correlation over a wide pressure range for anvils of different shapes. The data are consistent with each other in different runs and with previously reported data obtained below ~300 GPa(*9, 25, 28*) (Fig. 2). We fitted this combined data set over the entire pressure range of 0–477 GPa using equation(*28*), $P = A \cdot \frac{\Delta\omega}{\omega_0} + B \cdot \left(\frac{\Delta\omega}{\omega_0}\right)^2$, where $\Delta\omega$ is the shift of the diamond Raman edge of the stressed diamond anvil, $\omega_0 = 1332.5$ cm$^{-1}$ is the initial position of the diamond Raman edge of unstressed diamond anvils at ambient pressure, and A and B are the fitting parameters. The refined universal parameters A = 517±5 and B = 764±14 fit the data well and apply for accurate pressure estimation from ambient pressure to 477 GPa.

The new diamond scale contradicts the currently used diamond scale(*29*) at pressures above 250 GPa, reaching a difference in pressure values over 100 GPa at approximately 500 GPa (Fig. 2). The highest pressure of ~477 GPa based on a gold equation of state(*23*) measured in the present study corresponds to the record highest shift of the diamond Raman edge of stressed anvils of ~2026 cm$^{-1}$ (592 GPa according to the scale(*29*)). It was achieved in run 1 using toroidal diamond anvils with culets of diameter ~10 μm. Note that a similar pressure of ~446 GPa (the diamond Raman edge at ~1996 cm$^{-1}$) was reached in run 3 with traditional, not toroidal, but double-beveled diamond anvils (culet 12 μm).

The diamond edge Raman scale was robust to different load distribution patterns on the diamond tip. It works well for diamond anvils of different shapes (see Fig. S1 in Supplementary Materials). Notably, the refined diamond scale is valid for pressure estimation at the center of the diamond anvils and for a large area of ~20-30 μm around the diamond tip (see Fig. 1D and Fig. S2-S4 in the Supplementary Materials).

The accuracy of the pressure evaluation from the X-ray diffraction data is affected by the deviatoric stress α(σ$_3$ − σ$_1$) (the difference between the maximal and minimal eigenvalues of the stress tensor) in the Au sample (Fig. 1F). The values of α(σ$_3$ − σ$_1$) reached ~10–15 GPa at the highest pressures, which was lower than the reference data from the ramp experiments(*23*). The lattice volume of gold was determined from the X-ray diffraction data using the (111) diffraction peak position, which is the least affected by non-hydrostatic compression(*2*).

The universal scale for a sample pressure estimation is crucial for collating experimental results from different groups and comparison with theoretical predictions. Because of the inconsistency in the pressure scales used in the reported data, the intriguing observations of pressure-induced phase transitions in hydrogen and its metallization are difficult to compare in the multi-megabar pressure range. There is no consensus on which pressure scale should be used. The latest diamond edge Raman scale calibration up to 410 GPa(*29*) and more conservative scale(*28*) have both been used(*12, 31*) to study hydrogen, but they differ dramatically above ~300 GPa. This ambiguity raises the question of how suitable the diamond edge scale will be for hydrogen. The diamond scales are based on the equation of state of metals, but hydrogen is much softer. In addition, the compression of hydrogen is sensitive to the thickness and diameter of the sample and material of the gasket because a particular arrangement of experiments influences the stresses in the diamond anvils and the Raman spectrum. In particular, it develops in a pronounced sharp peak at the high-wavenumber Raman edge of diamond anvils recorded over the hydrogen sample (Fig.S5, run 7), whereas it manifests as a weak step in Raman spectra recorded over gold samples (Fig.S5).

We checked the validity of the diamond edge Raman scale by measuring the lattice volume of gold placed inside the hydrogen sample in run 7. Gold is the only metal that likely does not react with hydrogen. Theoretical works are contradictory: Kim et al.(*32*) predicted the formation of AuH above 220 GPa, whereas Gao et al.(*33*) demonstrated that AuH is thermodynamically unfavorable against elements. Experimentally, no hydrides of gold have been found at pressures up to 113 GPa and temperatures up to 600 K(*34*). In the present experiments performed at room temperature, gold did not form a hydride up to the highest pressure of ~290 GPa, as both the face-centered cubic crystal structure and the lattice volume of the sample corresponded to pure gold(*2, 3*). The pressure estimated from the gold sample surrounded by hydrogen in run 7 is in good agreement with the pressure value simultaneously estimated from the new diamond Raman edge scale, which strongly supports the application of the updated diamond edge Raman scale to hydrogen samples.

The position of the hydrogen vibron in the Raman spectra is sensitive to pressure. For example, one can observe a large scatter in the data of ~20 GPa from run to run on the plots of the pressure dependence of hydrogen vibron (*9, 35*). The accuracy of the pressure determination can be significantly improved by using hydrogen itself as a pressure gauge by calibrating the pressure-induced shift of hydrogen vibron $\omega(P)$ against the gold chip placed in hydrogen. We found that hydrogen vibrons at ~2990 cm$^{-1}$ and ~4085 cm$^{-1}$ correspond to a pressure of ~283 GPa, as determined from the gold chip (Fig. 3A inset). At higher pressures, diamond anvils break during exposure to intense synchrotron X-rays; this premature failure is common(*36*). This calibration of the hydrogen vibron can serve as a reference, and the pressure correction can be performed by shifting the whole plot of $\omega(P)$ to the calibrated point at $P$=283 GPa, as shown in Fig. 3A.

The phase diagram of hydrogen was corrected according to the updated universal diamond scale (Fig. 3C), in particular, the metallization pressure was corrected to 300 GPa. Above this pressure, as follows from the electrical measurements(*9-11*), hydrogen transforms into a semimetal through a semiconducting state. This transformation is consistent with the current theoretical predictions of the metallization of hydrogen through the closing of the indirect gap(*37-39*). The direct gap is opened in the semi-metallic state, and hydrogen weakly absorbs light, allowing Raman(*10*) and IR(*12*) spectroscopy. At higher pressures, a sharp decrease in IR absorption was reported at ~427 GPa(*12*), and the disappearance of the Raman signal at ~450 GPa(*10*). According to the new scale, both observed transitions are, in fact, at close pressures of ~400-410 GPa (Fig. 3B), indicating that both observations are likely related to the same transition. Experimentally, the nature of this transition is difficult to reveal because X-ray diffraction or spectroscopic data are not available above this pressure. Theoretically, a structural transformation from *C2/c* to *Cmca*-12 molecular phases was predicted at 423 GPa before the atomic Cs-IV phase at 447 GPa. In another scenario, the direct band of hydrogen closes at ~450 GPa(*37, 38*), or ~430 GPa(*39*) in the same molecular phase *C2/c*-24. Further studies are required to converge the experimental and theoretical data.

Correct pressure evaluation is crucial for the interplay between theory and experiment, which can be very fruitful, as illustrated by recent progress in near-room temperature superconductivity. The experiment, which is often guided by calculations, provides discrepancies that improve the computational models and fundamental understanding. Reliable predictions of crystal structures and properties of matter at high pressures, together with the development of the synthesis of new compounds, will advance science and technology.

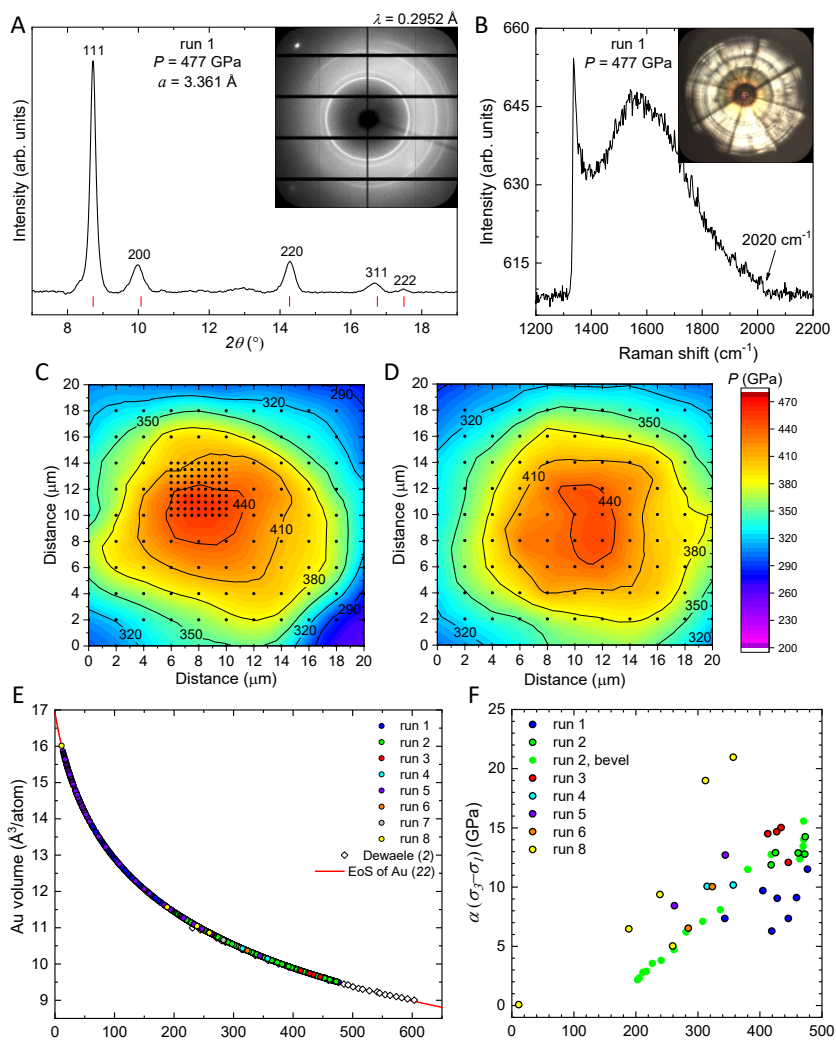

**Figure 1.** X-ray diffraction and Raman measurements for the diamond edge Raman scale. (A) X-ray powder diffraction pattern of gold in run 1 at the maximum pressure value of 477 GPa. Red ticks indicate the calculated positions of gold. The inset shows the original diffraction pattern. (B) Raman spectrum of the stressed diamond anvil in the same run demonstrates the position of the diamond Raman edge at ~2020 cm$^{-1}$. The inset shows the photo of the sample. (C and D) Spatial distribution of pressure on the diamond tip in run 2 reconstructed from the X-ray powder diffraction data using a gold equation of state(*23*) (C) and from Raman spectroscopy data of the stressed diamond anvil (D). Black points are spots of measurements. The pressure values estimated by two different techniques agree well. (E) Lattice volume of gold measured up to 477 GPa in runs 1-8 and estimated based on the position of the (111) diffraction peak. Pressure values were estimated from the isotherm obtained from the ramp compression(*23*). (F) Uniaxial stress sustained by gold samples in runs 1-6 and 8. Circles with black edges correspond to values of uniaxial stress calculated at the highest pressure in each compression cycle. Plain green circles correspond to values of uniaxial stress in the gold sample in run 2 at different spots by moving from the center of diamond culet to the bevel (the total length of ~39.6 μm, the step of ~2.8 μm).

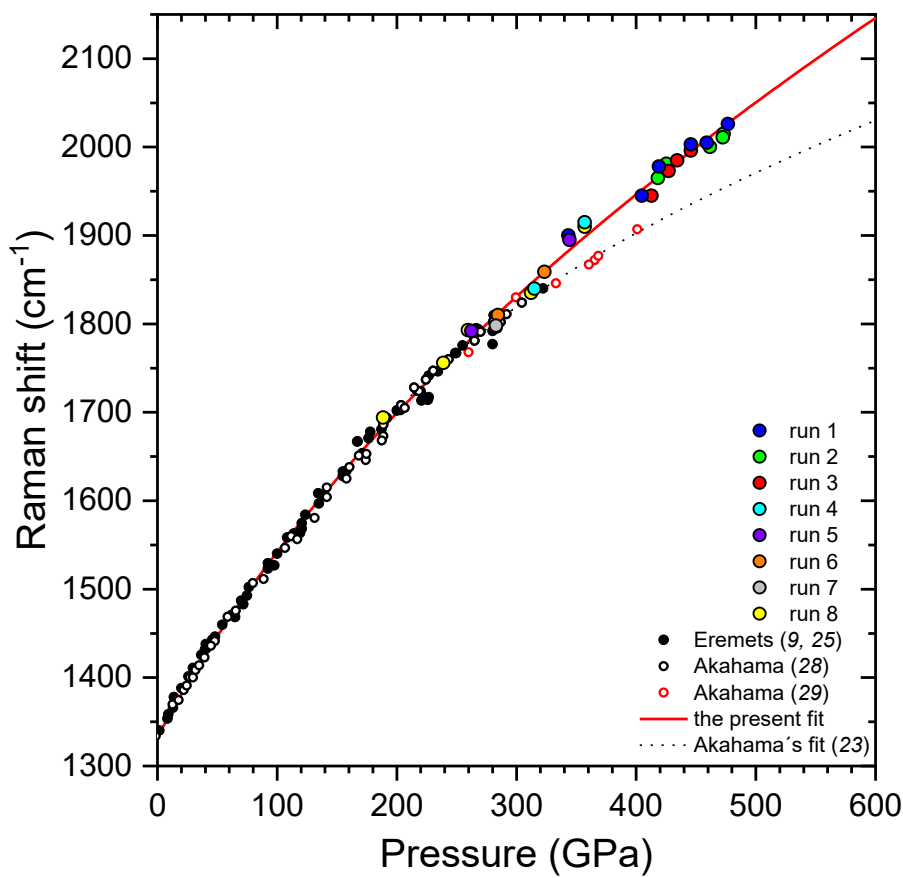

**Figure 2.** The universal diamond edge Raman scale. Large colorful circles correspond to the pressure dependence of the high-wavenumber Raman edge of the stressed diamond anvil measured in runs 1-8 of the present study. Small black circles are the data from Ref.(*9, 25*) (pressures values were estimated using the ruby scale(*24, 25*) below ~200 GPa and equation of state of Au(*40*) above ~200 GPa). Open black and red circles correspond to the data from Ref.(*28, 29*) after correcting the pressure values using the recent ramp compression data of Pt(*23*). The solid red curve fits the combined experimental data set, including the present and previous data measured below ~300 GPa(*9, 25, 28*). The black dotted curve is the updated diamond edge Raman scale of Akahama(*23*) for the high-pressure range above 200 GPa.

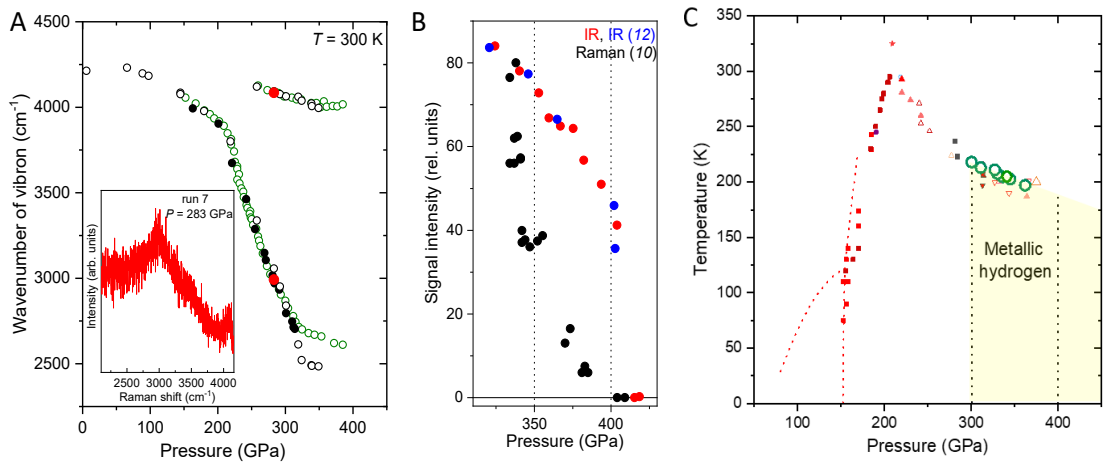

**Figure 3.** Comparison of different experimental studies of hydrogen at high pressures. (A) Pressure dependence of the hydrogen vibrons at room temperature. Solid black, open black, and open grey circles correspond to data from Ref (*9*). Open green points are from Ref (*31*); in this work, the pressure scale(*28*) was used, which is close to the present scale in this pressure range. The pressure for the rest was determined from the present diamond edge scale. Red circles are the data measured in run 7 of the present study. The open circles were shifted to lower pressures by 17 GPa to be consistent with the wavenumber of hydrogen vibron at the calibrated point (red circles). (B) Comparison of the infrared absorption data(*12*) (blue circles) and the Raman data(*10*). Pressure values for infrared and Raman data were recalculated according to the present scale (the values were shifted by ~10 GPa to lower pressures for infrared measurements(*12*); the maximum pressure of 475 GPa was significantly lowered to 409 GPa for Raman data(*10*). The comparable updated pressure values for the infrared and Raman data likely indicate the same transition. (C) The updated phases diagram of hydrogen according to the present pressure scale. The color points are from Ref. (*10*), green circles are from Ref. (*11*). Vertical dotted black lines indicate the boundary of the metallic state of hydrogen at 300 GPa (closure of the indirect gap) and a possible structural or electronic transition at 400 GPa, which likely corresponds to closure of the direct bandgap or a structural transition.

## ACKNOWLEDGMENTS

M.I. E. is thankful to the Max Planck community for the support, and Prof. Dr. U. Pöschl for the constant encouragement. X-ray diffraction were performed at GeoSoilEnviro CARS (The University of Chicago, Sector 13), Advanced Photon Source (APS), Argonne National Laboratory (runs 1-7) and DESY (Hamburg, Germany), a member of the Helmholtz Association HGF (run 8). GeoSoilEnviro CARS is supported by the National Science Foundation-Earth Sciences (EAR-1634415) and Department of Energy-GeoSciences (DE-FG02-94ER14466). This research used resources of the Advanced Photon Source, a U.S. Department of Energy (DOE) Office of Science User Facility operated for the DOE Office of Science by Argonne National Laboratory under Contract No. DE-AC02-06CH11357. Parts of this research were carried out at PETRA-III using P02.2. We appreciate help of Dr. I. A. Troyan and Dr. H. P. Liermann with the X-ray diffraction experiments at PETRA-III.

## AUTHOR CONTRIBUTIONS

M.I.E. and V.S.M equally contributed to the manuscript. M.I.E. supervised the work. M.I.E., V.S.M. and A.P.D. prepared the samples. V.B.P. and S.C. performed X-ray diffraction and Raman measurements at APS. V.S.M. processed the data. M.I.E. and V.S.M. wrote the manuscript.

## COMPETING INTERESTS

The authors declare no competing interests.

Supplementary Materials for
# Universal diamond edge Raman scale to 0.5 terapascal. The implication to metallization of hydrogen.

M. I. Eremets[1]*, V. S. Minkov[1], P. P. Kong[1], A. P. Drozdov[1], S. Chariton[2], V. B. Prakapenka[2]

[1]*Max Planck Institute for Chemistry; Hahn Meitner Weg 1, Mainz, 55128, Germany*
[2]*Center for Advanced Radiation Sources, University of Chicago; 5640 South Ellis Avenue, Chicago, Illinois, 60637, USA*

*Corresponding author. Email: m.eremets@mpic.de

**Materials and Methods**
**Supplementary figures S1 to S8**
**References**

## MATERIALS AND METHODS

**Diamond anvil cell preparation.**

High pressure was generated using DACs with a diameter of 25 mm and a length of 35 mm. The load is applied by pushing the piston, which is moved by screws at the top of the DAC. Anvils are made of synthetic or natural diamonds with culets that are normal to the [001] crystallographic direction. They were beveled at ~8° to a diameter of ~250-400 μm. We used small culets of ~10-20 μm with different shapes of the diamond tip to achieve the highest pressures (Fig. S1). The toroidal shape of the diamond anvils was made with the aid of a focused beam of xenon ions (FERA3, Tescan). The ion beam current was set between 1 and 10 nA under an accelerating voltage of 30 kV. The total duration of machining of the toroidal shape was approximately 1 hour per diamond anvil. The final profile of the diamond anvils was measured using a profilometer. For the sample loading, 250-μm-thick T301 stainless steel gaskets were preintended to a thickness of ~5 μm, and a hole with a diameter of ~20 mm was drilled by a laser. A piece of gold (SkySpring Nanomaterials, 99.99%) was placed in the prepared hole and clamped in the DACs. Gold was used as the standard for the pressure determination. In total, eight DACs were pressurized to ~200-300 GPa in the home laboratory and transferred to synchrotron facilities, where the pressure was further increased.

**X-ray diffraction and Raman measurements.**

X-ray diffraction data were collected at the beamlines 13-IDD at GSECARS, Advanced Photon Source ($\lambda$ = 0.2952 Å, a spot size of ~2.5×3.5 μm$^2$, Pilatus 1M CdTe detector) and P02.2 at PETRA III, DESY ($\lambda$ = 0.2910 Å, a spot size of ~3×3 μm$^2$, LAMBDA GaAs detector). The typical exposure time was varied between 1s and 5s. Reference samples of $LaB_6$ and $CeO_2$ were used to calibrate the distance between the sample and detector. The X-ray beam was cleaned using a pinhole to remove the beam wings. The processing and integration of the data and background subtraction were performed using the Dioptas software(*1*).

The Raman spectra were recorded using an advanced Raman setup at GESECARS(*2*). The Raman signals of stressed diamond anvils in contact with gold samples were excited by lasers with two different wavelengths: $\lambda_1$ = 660 nm and $\lambda_2$ = 532 nm. The switching from red to green laser helped to reduce and shift the strong luminescence of the stressed diamond, which may appear in synthetic diamonds at pressures above 300 GPa and overlap the diamond Raman edge signal of stressed anvils. The laser power of the incident light coming from a ×50 microscope objective (Mitutoyo) was reduced to ~3-5 mW to protect the diamond anvils from failure because of the enhanced absorption of the highly stressed anvils. The typical exposure time was 60-120 s per Raman spectrum.

At each pressure point, the position of the X-ray beam was visualized with the X-ray-induced luminescence of the sample, and the spot was aligned relative to the center of a diamond culet, which was observed under illumination. For the Raman measurements, the focused laser beam was aligned relative to the image of the anvil. This alignment was reproducible at the same stage as the fixed DAC was placed in the Raman and X-ray setups with an accuracy of ~1 μm. Thus, the X-ray and laser beams were aligned to probe the sample at the exact location. This is crucial because of the large pressure gradients over the diamond tip. The original X-ray powder diffraction patterns of gold and the corresponding Raman spectra of stressed diamond anvils measured at different pressures in different runs are illustrated in Fig. S5.

To measure the spatial pressure distribution at the diamond anvil tip, we performed X-ray diffraction and Raman mapping on an area of 20×20 μm$^2$ with a horizontal and vertical step of 2 μm. At certain highest pressure points, the grid step was reduced to 0.5–1 μm. In addition, at selected pressure points, diamond anvils were scanned horizontally and vertically up to 70 μm away from the center of the diamond culet (Fig. S2-S4 and S6).

**Pressure determination**

The diamond Raman edge was determined as a point at half the height of the step at the high-wavenumber edge in the Raman spectra of the stressed diamond anvil (Fig. 1B). We assigned this wavenumber to the pressure value estimated from the X-ray diffraction measurements by referring the refined lattice volume of the gold sample to the reduced isotherm obtained from the ramp experiments (*3*).

We also estimated how the deviatoric stresses $\alpha(\sigma_3 - \sigma_1)$ (the difference between the maximal and minimal eigenvalues of the stress tensor) in our Au samples could influence the accuracy of the pressure determination. The deviatoric stress was estimated by applying the commonly used method (*4-6*). This method is based on the analysis of the positions of the diffraction peaks because different peaks shift differently under non-hydrostatic stress. In a cubic crystal lattice, all diffraction peaks yield identical lattice parameters. In contrast, in the strained crystal, they differ by a factor proportional to $\sigma_3 - \sigma_1$ and the elastic anisotropy of the crystal. The single-crystal elastic constants were taken from the low-pressure measurements in Ref. (*7*) and extrapolated to our high-pressure values. Five diffraction peaks of gold (111), (200), (220), (311), and (222), which were measured in most of our runs, were used for the analysis. We used an iso-stress assumption in our powder samples (α, which varies from 0.5 to 1, was set as 1), which provides minimum values of $\sigma_3 - \sigma_1$. The deviatoric stresses reach ~10-15 GPa, i.e. ~2-3% at 500 GPa (Fig. 1F), and are less than those sustained in the ramp experiments(*3*).

# SUPPLEMENTARY FIGURES

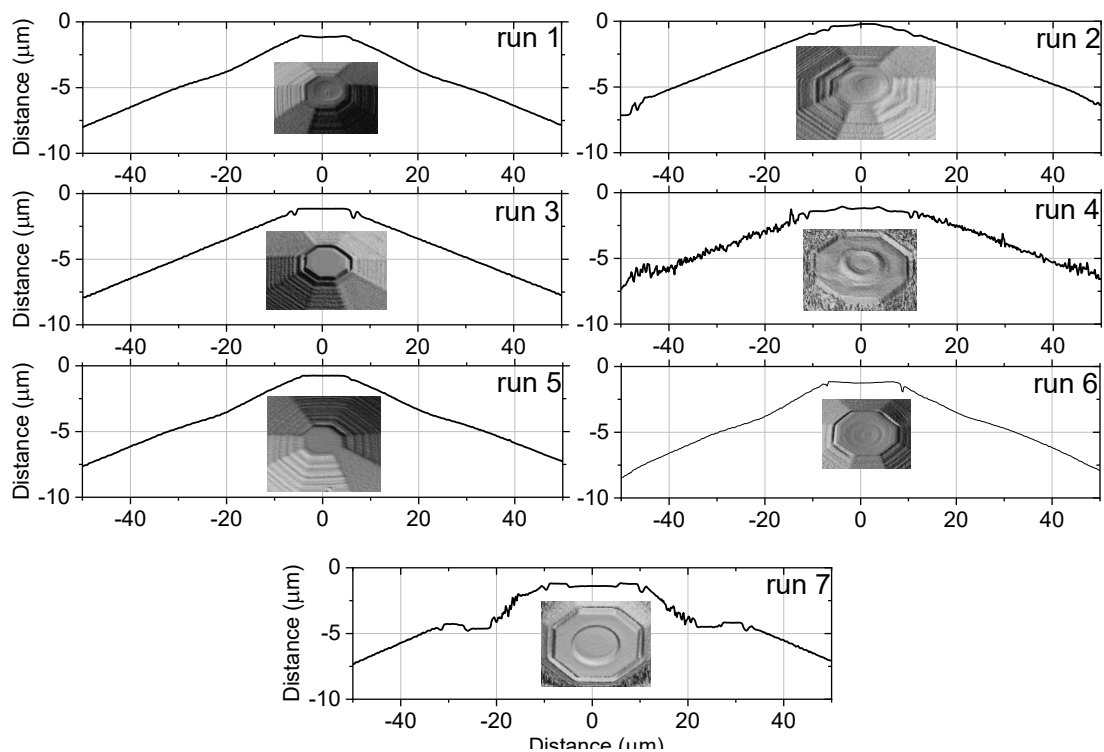

**Supplementary figure S1.** Profiles of diamond anvil tips used in runs 1-7 of the present study. Insets illustrate the geometry of diamond culets. In run 8 we used anvils with culet of diameter 8 μm beveled at 7.5° to diameter 320 μm.

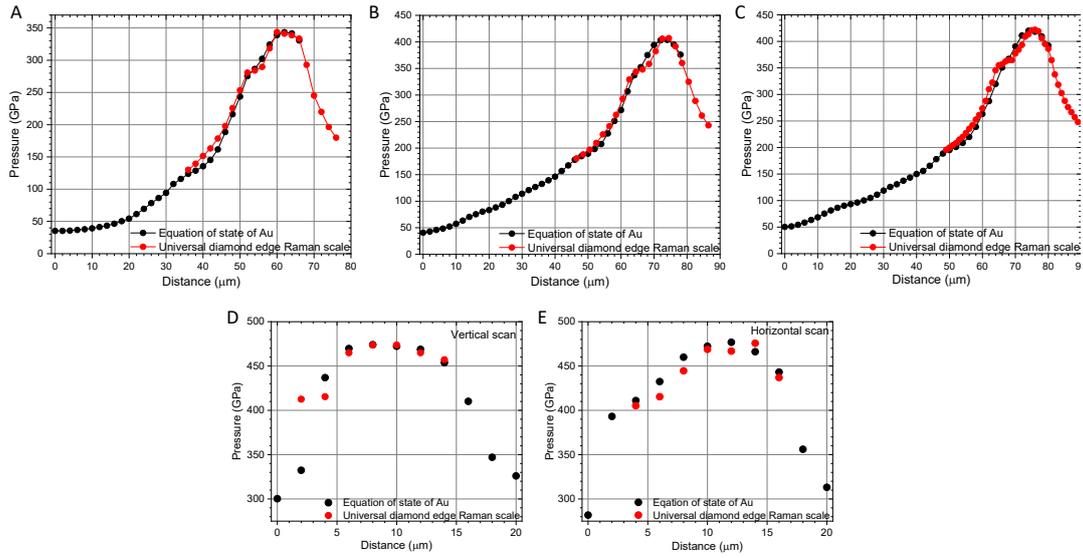

**Supplementary figure S2.** Distribution of pressure on the diamond tip in run 1 at several loads. (A-C) Extended pressure distribution plots reconstructed from one-dimensional X-ray diffraction (black circles) and Raman (red circles) mappings. (D and E) Plots of pressure distribution on the diamond culet at the highest load. Pressure values were estimated using the equation of state of Au(*3*) (X-ray diffraction) and the present universal diamond edge Raman scale.

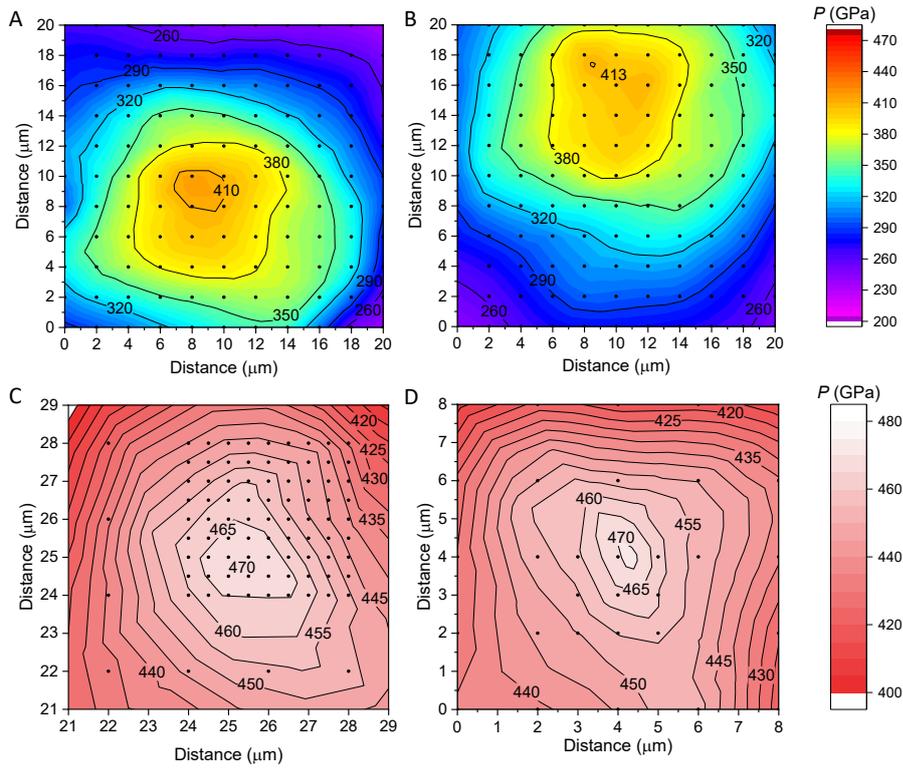

**Supplementary figure S3.** Spatial distribution of pressure on the diamond tip in run 2 at several loads. (A and B) Extended pressure distribution plots reconstructed from two-dimensional X-ray diffraction and Raman mappings. (C and D) Plots of pressure distribution on the diamond culet at the highest load. Pressure values were estimated using the equation of state of Au(*3*) (X-ray diffraction, panels A and C) and the universal diamond edge Raman scale (panels B and D). Black points are spots of measurements.

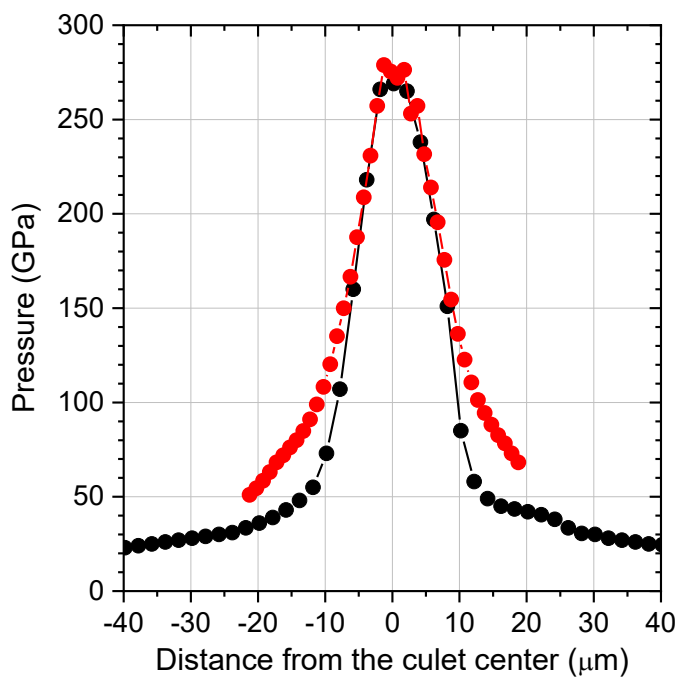

**Supplementary figure S4.** Distribution of pressure on the diamond tip (culet 10 μm) in run 5. The plot is reconstructed from one-dimensional X-ray diffraction (black circles) and Raman (red circles) mappings. Pressure values were estimated using the equation of state of Au(*3*) (X-ray diffraction) and the universal diamond edge Raman scale. The difference in pressures values at distances farther ~10 μm from the diamond culet center is naturally expected due to very ununiformed stresses distribution in the stressed anvil(*8*). Importantly that the pressure values well coincide within the diamond culet area. This fact supports the universality of the present pressure scale.

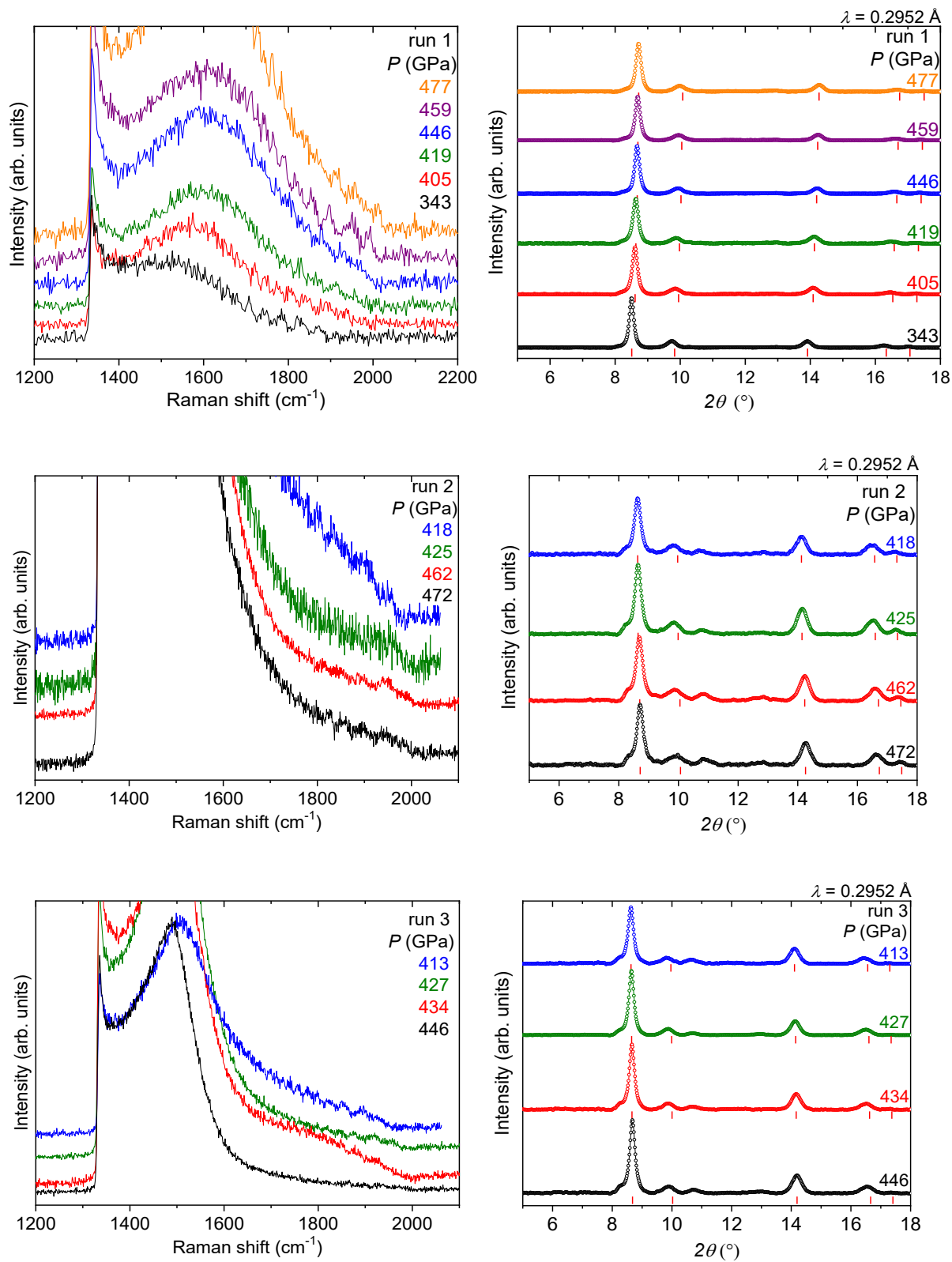

**Supplementary figure S5.** Original Raman spectra of stressed diamond anvils (left) and corresponding X-ray diffraction powder patterns of gold samples (right) measured in runs 1-8.

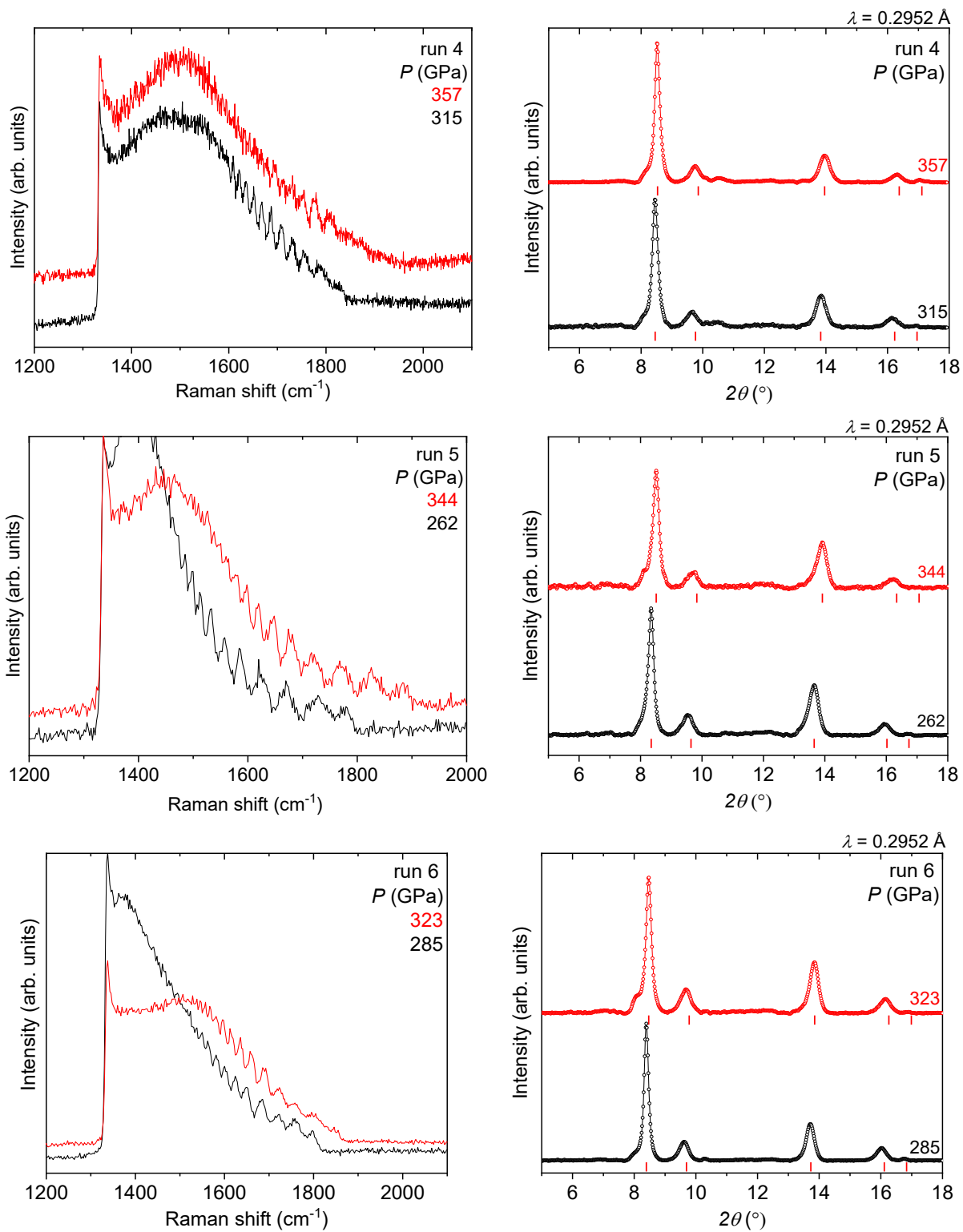

**Supplementary figure S5 (continuation).** Original Raman spectra of stressed diamond anvils (left) and corresponding X-ray diffraction powder patterns of gold samples (right) measured in runs 1-8.

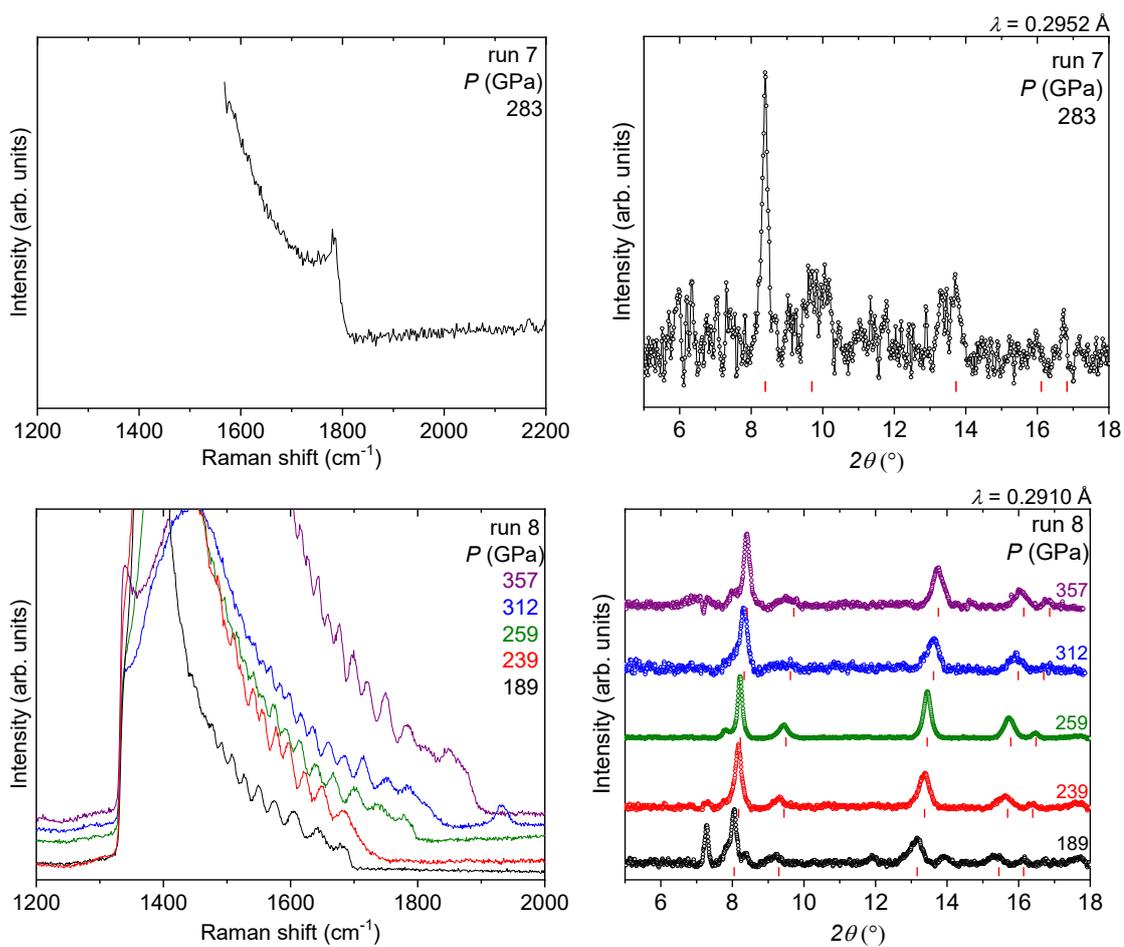

**Supplementary figure S5 (continuation).** Original Raman spectra of stressed diamond anvils (left) and corresponding X-ray diffraction powder patterns of gold samples (right) measured in runs 1-8.

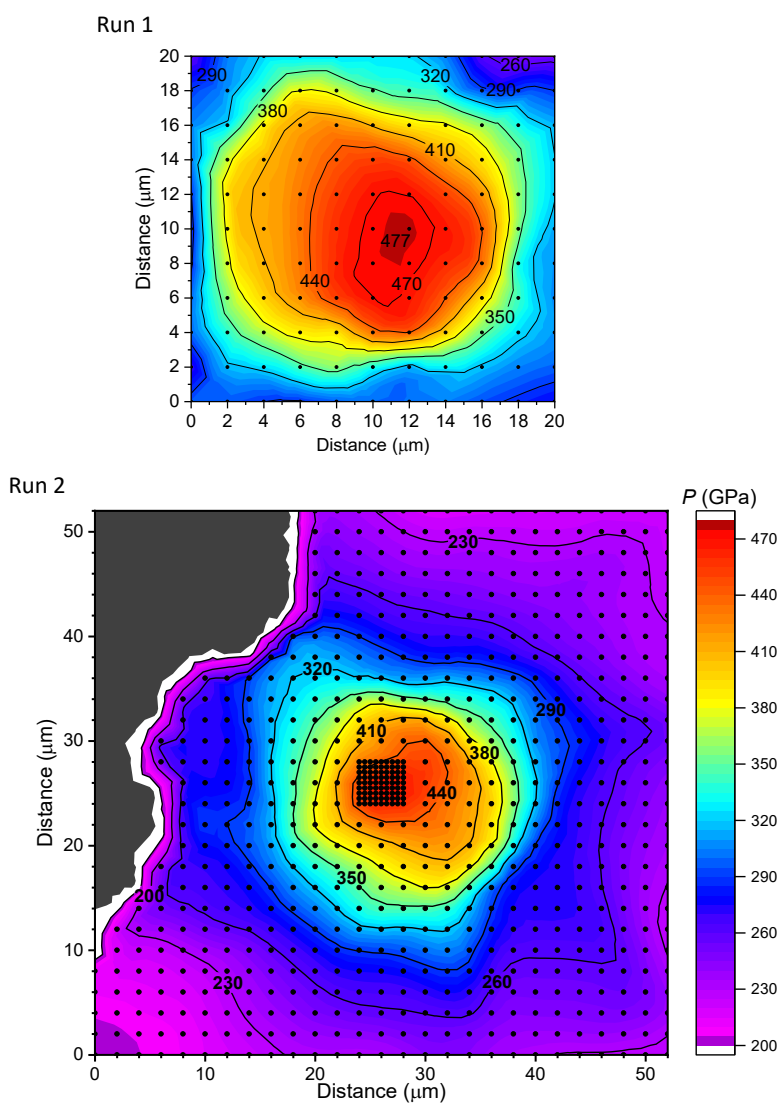

**Supplementary figure S6.** Spatial distribution of pressure on the diamond tip in runs 1 and 2 at maximum loads. The plots are reconstructed from two-dimensional X-ray diffraction mapping. Black points are spots of measurements.

Run1

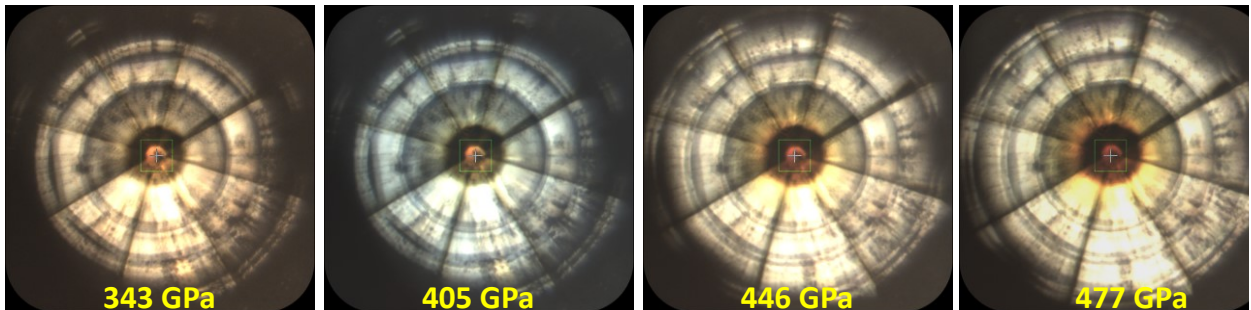

Run 2

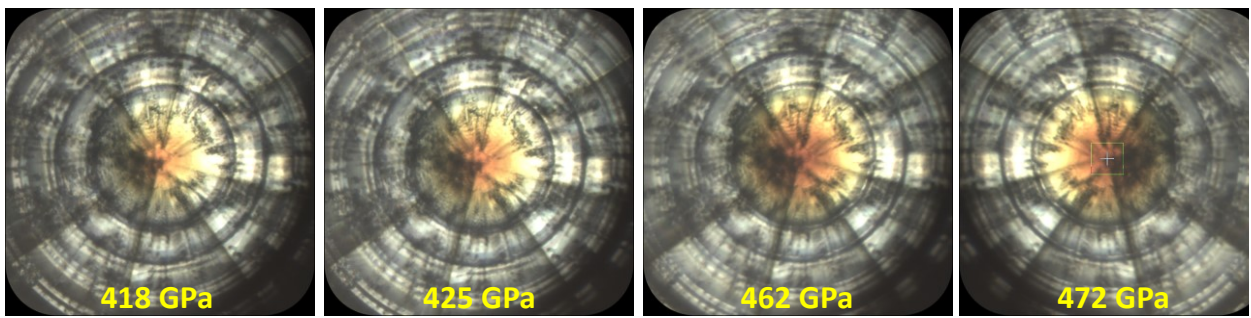

Run 3

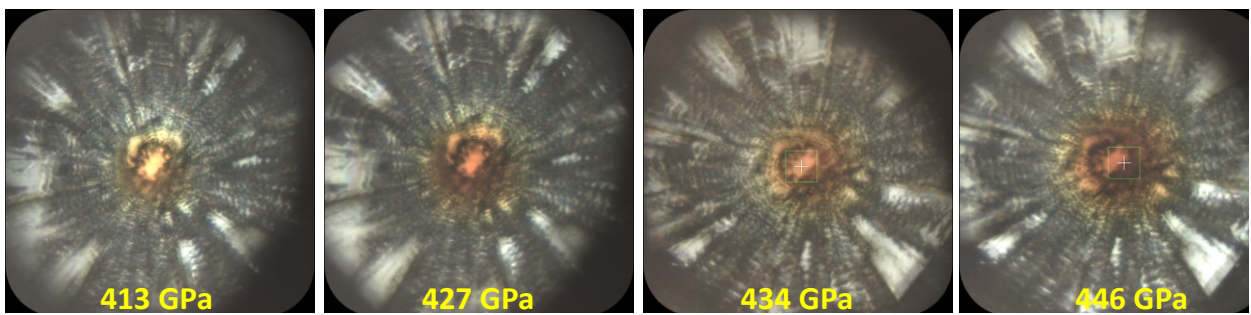

**Supplementary figure S7.** Photos of diamond anvils with gold samples at highest achieved pressures in runs 1-3. Diamond anvils become darker with pressure.

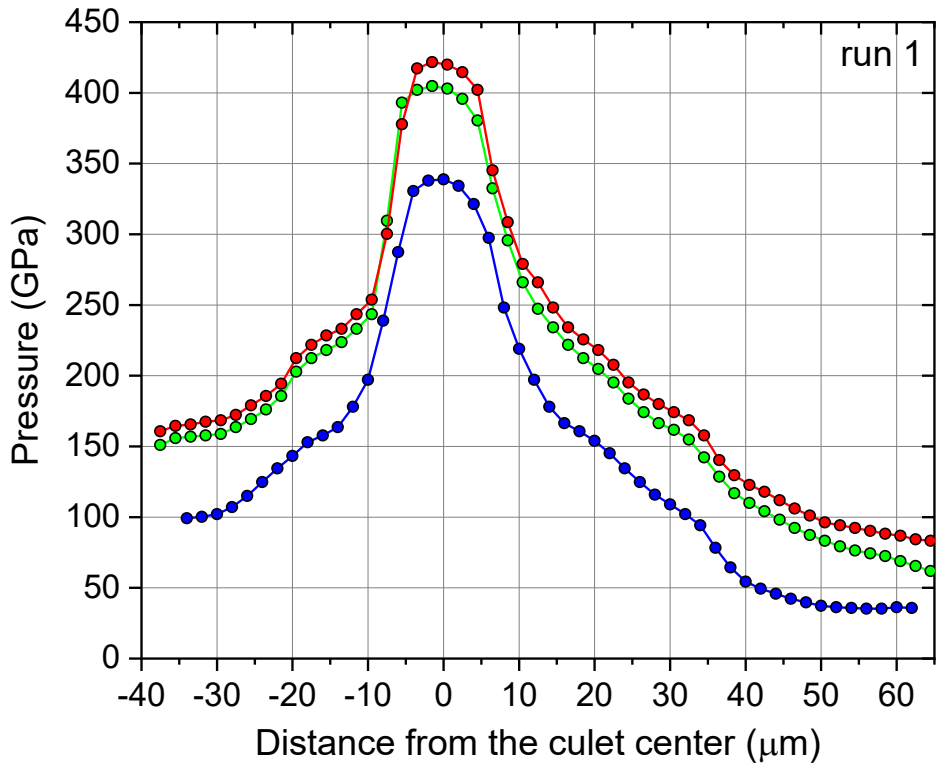

**Supplementary figure S8.** Distribution of pressure on the diamond tip in run 1 at several loads. The plot is reconstructed from one-dimensional X-ray diffraction mapping. Pressure values were estimated using the equation of state of Au(*3*).